\documentclass{ws-ijmpa}
\usepackage{amssymb,graphicx,amsmath,graphics,xfrac,url,subfigure,booktabs,adjustbox}
\usepackage[super,compress]{cite}

\begin{document}
	
\title{Bound states of a Klein-Gordon particle in presence of a smooth potential well}
	
\author{Eduardo L\'opez}
\address{School of Physical Sciences and Nanotechnology, Yachay Tech University, 100119 Urcuqu\'i, Ecuador}
	
\author{Clara Rojas \thanks{crojas@yachaytech.edu.ec}}
\address{School of Physical Sciences and Nanotechnology, Yachay Tech University, 100119 Urcuqu\'i, Ecuador}

\maketitle
	
\begin{history}
\received{\today}
\end{history}
	
\begin{abstract}
We solve the one-dimensional time-independent Klein-Gordon equation in presence of a  smooth potential well.  The bound state solutions are given in terms of the  Whittaker $M_{\kappa,\mu}(x)$ function, and the antiparticle bound state is discussed in terms of potential parameters.
		
\keywords{Hypergeometric functions, Klein-Gordon equation, bound-state solutions}
\end{abstract}
	
\ccode{PACS numbers: 02.30.Gp,03.65.Pm, 03.65.Nk}

\section{Introduction}

The discussion of the overcritial behavior of bosons requires a full understanding of the single particle spectrum. For short range potentials, the solutions of the Klein-Gordon equation can exhibit spontaneous production of antiparticles as the strength of an external potential reaches certain value $V_0$ \cite{rafelski:1978}.
In $1940$, Schiff, Snyder and Weinberg \cite{schiff:1940} carried out one of the earliest investigations of the solution of the Klein-Gordon equation with a strong external potential. They solved the problem of the square well potential and discovered that there is a critical point $V_{cr}$ where the bound antiparticle mode appears to coalesce with the bound particle. In $1979$, Bawin \cite{bawin:1979} showed that the antiparticle p-wave bound state arises for some conditions on the potential parameters. 

In the present article, we solve the one-dimensional time-independent Klein-Gordon equation for a
smooth potential well.  This smooth potential well is a short-range potential which support antiparticle bound states and we can  determine how the shape of the potential affects the pair creation mechanism. This potential is interesting because varying the smoothness of the curve can be represented from the square potential well to the cusp potential well. We  show that the antiparticle bound states arise in the limit  of the square potential well, and  in the limit of a cusp potential well.

The Klein-Gordon equation is used to describe spin-0 particles. 
The analytical solution of the one-dimensional time-independent Klein-Gordon equation for different potentials has been caused of a lot of interest in recent years, for both bound states \cite{rojas:2006a, rojas:2006b,lutfuoglu:2012,alpdogan:2013,hassanabadi:2014,di:2016,lutfuoglu:2016} and scattering solutions \cite{rojas:2005,rojas:2007,villalba:2007,lutfuoglu:2012,rojas:2014a, rojas:2014b,hassanabadi:2014,ikot:2015,lutfuoglu:2016,chabab:2016,lutfuoglu:2018,aquino:2019,rojas:2020a}. It has allowed the understanding of several physical phenomena of Relativistic Quantum Mechanics such as the antiparticle bound state \cite{schiff:1940,bawin:1974}, transmission resonances \cite{rojas:2005,rojas:2007,villalba:2007}, and superradiance  \cite{manogue:1988,rojas:2014a,molgado:2018,rojas:2019a}.

The article is structured as follows: Sec. 2 is devoted to discuss
the Klein-Gordon equation. In Sec. 3 we present the smooth potential well. In Sec. 4 we solve the Klein-Gordon
equation in the presence of a one-dimensional smooth potential well. In Sec. 5 we derive the equation governing the eigenvalues corresponding to the bound states and compute the bound states. We also show the
dependence of supercritical states on the strength and shape of the
potential. Finally, in Sec. 6, we briefly summarize our results.

\section{The Klein-Gordon equation}

The Klein-Gordon equation for free particles, in natural units $\hbar=c=m=1$, is given by \cite{greiner:1987},

\begin{equation}
\label{KG}
\hat{p}^\mu\hat{p}_\mu \varphi = \varphi,
\end{equation}
being $\hat{p}^\mu=i\left(\dfrac{\partial}{\partial t},-\vec{\nabla}\right)$, Eq. \eqref{KG} becomes:

\begin{equation}
\left(\Box+1\right)\varphi=0.
\end{equation}

We need to solve the Klein-Gordon equation interacting with a spatially one-dimensional potential, then we start finding the form of the Klein-Gordon equation with the interaction of an electromagnetic field.

\bigskip
The electromagnetic field is described by the four-vector \cite{greiner:1987}:

\begin{eqnarray}
A^\mu&=&(A_0,\vec{A}), \\
A_\mu&=&\eta_{\mu\nu} A^\nu=(A_0,-\vec{A}),
\end{eqnarray}
where $\eta_{\mu\nu}=\textnormal{diag}(1,-1)$.

\bigskip
The minimal coupling of the electromagnetic field is expressed in the form,

\begin{eqnarray}
\hat{A}^\mu&\rightarrow&\hat{p}^\mu-e A^\mu,\\
\hat{A}_\mu&\rightarrow&\hat{p}_\mu-e A_\mu .
\end{eqnarray}

\bigskip
The one-dimensional Klein-Gordon equation  minimally coupled to a vector potential $A^{\mu}$
can be written as \cite{greiner:1987}:

\begin{equation}
\label{KG_gen}
(\hat{p}^\mu- e A^\mu)(\hat{p}_\mu- e A_\mu)\varphi=\varphi.
\end{equation}

\medskip
Consider a spatially one-dimensional potential $e A_0=V(x)$, $\vec{A}=0$, and a stationary solution of the Klein-Gordon equation  $\varphi(x,t)=\phi(x)e^{-i E t}$, Eq. \eqref{KG_gen} can be written as:

\begin{equation}
\label{KG_x}
\frac{\mathrm{d}^{2} \phi(x)}{\mathrm{d}x^{2}}+\left\{[E-V(x)]^{2}-1\right\}\phi(x)=0,
\end{equation}
where $E$ is the energy of the particle.

\section{The smooth potential well}

\begin{figure}[htbp]
\begin{center}
\includegraphics[scale=0.5]{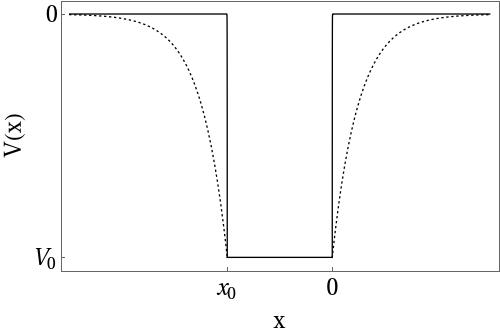}
\end{center}
\caption{\label{fig_pot}{Smooth  potential well for $V_0=2$ with $a=0.5$ (dotted line) and square potential well for $a=0.001$ (solid line).}}
\end{figure}

The smooth potential well is given by:

\begin{equation}
\label{potential}
V(x) =
\left\{
\begin{array}{ll}
\quad	- V_0 e^{\sfrac{(x-x_0)}{a}}, \quad \textnormal{for}  \quad x<x _0,\\
\quad - V_0, \quad \textnormal{for} \quad x_0 \leq x\leq 0,\\
\quad	- V_0 e^{-\sfrac{x}{a}}, \quad \textnormal{for} \quad  x > 0,
\end{array}
\right.
\end{equation}
where $V_0$ represents the depth of the potential well, $a$ gives the smoothness of the curve, and $x_0$ represents the widht of the smooth potential well. The form of the potential \eqref{potential} is showed in the Fig. (\ref{fig_pot}). From Fig. \ref{fig_pot} we can note that  the smooth potential well reduces to the square potential well considering $a \rightarrow 0$. Also this potential reduces to the cusp well for $a > 0$ and $x_0=0$.

\section{Bound State Solutions}

\subsection{Bound state solutions for $x<x_0$}

The bound state solutions for $x<x_0$ are obtained by solving the differential equation

\begin{equation}
\label{eq_x_I}
\frac{\mathrm{d}^2\phi_{\textnormal{I}}(x)}{\mathrm{d}x^2}+\left\{\left[E+V_0e^{\frac{(x-x_0)}{a}}\right]^2-1\right\}\phi_{\textnormal{I}}(x)=0.
\end{equation}

\medskip
On making the change of variable $y=2iaV_0e^{\frac{(x-x_0)}{a}}$, Eq. (\ref{eq_x_I}) becomes

\begin{equation}
\label{eq_y_I}
y\frac{\mathrm{d}}{\mathrm{d}y}\left(y\frac{\phi_{\textnormal{I}}}{\mathrm{d}y}\right)-\left[\left(iaE+\frac{y}{2}\right)^2+a^2\right]\phi_{\textnormal{I}}=0.
\end{equation}

Putting $\phi_{\textnormal{I}}=y^{-\sfrac{1}{2}} f(y)$ we obtain the Whittaker differential equation

\begin{equation}
\label{f}
f(y)''+\left[-\frac 1 4 ´-\frac{iaE}{y} +\frac{\sfrac{1}{4}-a^2(1-E^2)}{y^2}\right]f(y)=0,
\end{equation}
which general solution is given by

\begin{equation}
\label{phi_y_I}
f(y)=c_1  M_{\kappa,\mu}(y)+d_1  W_{\kappa,\mu}(y),
\end{equation}
where $M_{\kappa,\mu}(y)$, $W_{\kappa,\mu}(y)$ are the Whittaker functions, $\kappa=-iaE$, and $\mu=a \sqrt{1-E^2}$. 
Then the solution of Eq. \eqref{eq_y_I} is given by,

\begin{equation}
\label{phi_y_II}
\phi_{\textnormal{I}}(y)=b_1 y^{-\sfrac{1}{2}} M_{\kappa,\mu}(y)+b_2 y^{-\sfrac{1}{2}} W_{\kappa,\mu}(y).
\end{equation}

\bigskip
Since we are looking for bounds states of equation \eqref{eq_x_I}, we choose to work with regular solutions of $\phi_{\textnormal{I}}$ along the $x$ axis. If $x \rightarrow - \infty, \quad y \rightarrow 0$, the limits of the Whittaker functions are given by

\begin{eqnarray}
\lim\limits_{y\rightarrow 0}\, y^{-\sfrac{1}{2}} M_{\kappa,\mu}(y) &\rightarrow& 0,\\
\lim\limits_{y\rightarrow 0} \,y^{-\sfrac{1}{2}} W_{\kappa,\mu}(y) &\rightarrow&  \textnormal{complex infinity}.
\end{eqnarray}

Then the regular solution is,

\begin{equation}
\label{phi_x_I_regular}
\phi_{\textnormal{I}}(x)=b_1 (2 i a V_0)^{-\sfrac{1}{2}} e^{-\frac{(x-x_0)}{2a}} M_{\kappa,\mu}\left[2 i a V_0 e^{\frac{(x-x_0)}{a}}\right].
\end{equation}

\bigskip
\subsection{Bound state solutions for $x_0\leq x\leq 0$} 

The bound solutions for $x_0<x<0$ are obtained by solving the differential equation

\begin{equation}
\label{x<x0and0}
\frac{\mathrm{d}^2\phi_{\textnormal{II}}(x)}{\mathrm{d}x^2}+\left[\left(E+V_0 \right)^2-1 \right]\phi_{\textnormal{II}}(x)=0.
\end{equation}

Eq. (\ref{x<x0and0}) has the general solution 

\begin{equation}
\label{phi_II}
\phi_{\textnormal{II}}(x)=b_3\,e^{-iqx}+b_4\,e^{iqx},
\end{equation}
where $q=\sqrt{\left(E+V_0\right)^2-1}$.

\bigskip
\subsection{Bound state solutions for $x >0$} 

The scattering solutions for $x>0$ are obtained by solving the differential equation

\begin{equation}
\label{eq_x_III}
\frac{\mathrm{d}^2\phi_{\textnormal{III}}(x)}{\mathrm{d}x^2}+\left[\left(E+V_0e^{-\frac{x}{a}}\right)^2-1\right]\phi_{\textnormal{III}}(x)=0.
\end{equation}

\medskip
On making the change of variable $z=2iaV_0e^{-\frac{x}{a}}$, Eq. (\ref{eq_x_III}) becomes

\begin{equation}
\label{eq_z_III}
z\frac{\mathrm{d}}{\mathrm{d}z}\left(z\frac{\phi_{\textnormal{III}}}{\mathrm{d}z}\right)-\left[\left(iaE+\frac{z}{2}\right)^2+a^2\right]\phi_{\textnormal{III}}=0.
\end{equation}

Putting $\phi_{\textnormal{III}}=z^{-\sfrac{1}{2}} g(z)$ we obtain the Whittaker differential equation

\begin{equation}
\label{g}
g(z)''+\left[-\frac 1 4 -\frac{iaE}{z} +\frac{\sfrac{1}{4}-a^2(1-E^2)}{z^2}\right]g(z)=0,
\end{equation}
which solution is given by

\begin{equation}
g(z)=b_5 M_{\kappa,\mu}(z)+b_6 W_{\kappa,\mu}(z).
\end{equation}

\bigskip
The solution of Eq. \eqref{eq_z_III} becomes

\begin{equation}
\label{phi_y_3}
\phi_{\textnormal{III}}(z)=b_5 z^{-\sfrac{1}{2}} M_{\kappa,\mu}(z)+b_6 z^{-\sfrac{1}{2}} W_{\kappa,\mu}(z).
\end{equation}

Searching the regular solution: if $x \rightarrow  \infty, \quad y \rightarrow 0$, the limits of the Whittaker functions are given by

\begin{eqnarray}
\lim\limits_{z\rightarrow 0}\, z^{-\sfrac{1}{2}} M_{\kappa,\mu}(z) &\rightarrow& 0,\\
\lim\limits_{z\rightarrow 0} \,z^{-\sfrac{1}{2}} W_{\kappa,\mu}(z) &\rightarrow&  \textnormal{complex infinity}.
\end{eqnarray}

Then, in terms of the variable $x$, the regular solution becomes

\begin{equation}
\label{phi_x_III_regular}
\phi_{\textnormal{III}}(x)=b_5 \,(2ia V_0)^{(-\sfrac{1}{2})} e^{\frac{x}{2a}} M_{\kappa,\mu}\left(2 i a V_0 e^{-\frac{x}{a}}\right).
\end{equation}

\section{Results and discussion}

In order to find bound states, we impose the condition that the  wave functions and their first derivatives must be continuous at $x=x_0$ and $x=0$.   Working algebraically with  the resulting system of equations we obtain that the energy eigenvalues must satisfy the equation

\begin{eqnarray}
\label{eigenvalues}
\nonumber
&&\dfrac{1}{4a^2} \left\{\left[1+2\kappa+2 i a(q-V_0)\right]M_{\kappa,\mu}(2 i a V_0)-(1+2 \kappa+2 \mu) M_{1+\kappa,\mu}(2 i a V_0)^2\right\}^2\\
\nonumber
&-&\dfrac{e^{-2 i x_0 q}}{4a^2} \left\{\left[1+2 \kappa -2 i a (q+V_0)\right] M_{\kappa ,\mu}(2 i a V_0)-(1+2 \kappa +2 \mu) M_{1+\kappa ,\mu}(2 i a V_0)\right\}^2=0.\\
\end{eqnarray}
 
\begin{figure}[!th]
\centering
\includegraphics[scale=0.5]{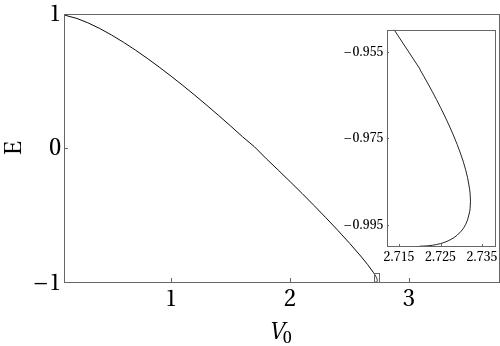}
\caption{Bound-state spectrum for $a=0.5$ and $x_0=-0.5$.}
\label{EvsV0_1}
\end{figure}

The explicit solutions of Eq. \eqref{eigenvalues}, showing the dependence of the energy $E$ on $V_0$, $a$ and $x_0$ can be determined numerically. Fig \eqref{EvsV0_1} shows the dependence of the spectrum of bound states on the potential strength $V_0$, for $  2.715 < V_0 \leq 2.735  $ two states appear, one with positive energy and another with negative energy. The bound antiparticle state  joins with the bound particle state at $V_\textnormal{cr}=2.7322$ where  they form a state with zero norm.

The normalization of the wave functions \eqref{phi_x_I_regular}, \eqref{phi_II}, and \eqref{phi_x_III_regular}, with the respective value of $b_1$, $b_3{\tiny }$, and $b_5$ in terms of $b_4$ obtained from the continuity conditions, is given by

\medskip
\begin{equation}
\label{norm}
N=2 \int^\infty_{-\infty}dx[E-V(x)]\phi(x)^*\phi(x).
\end{equation}

The norm of the Klein-Gordon equation  vanishes at $V_{cr}$, where both possible solution $E^{(+)}$ and  $E^{(-)}$ meet.
 Particle bound states ($E^{(+)}$) and antiparticle bound states ($E^{(-)}$) correspond to $N>0$ and $N < 0$ respectively. For $N=0$ both solutions meet and have the same energy. In Fig. \ref{EvsV0_1} the antiparticle bound state appears at $V_0=2.72$, and the norm of two points before reaching $V_\textnormal{cr}$ is showing in Table \ref{table_norm}.

\begin{table}[th!]
\begin{center}
\begin{tabular}{c|c|c|c}
\toprule
	$E$   & $V_0$   & $N$  & Comment \\
	\hline
- 0.979087   &   2.73      &     4.16463     & Particle bound state \\
- 0.996487   &    2.73     &  - 8.08205     & Antiparticle bound state\\
\bottomrule
\end{tabular}
\caption{Value of the norm for two points of the curve showed in Fig. \ref{EvsV0_1}.}
\label{table_norm}
\end{center}
\end{table}

\begin{figure}[!th]
\centering
\includegraphics[scale=0.5]{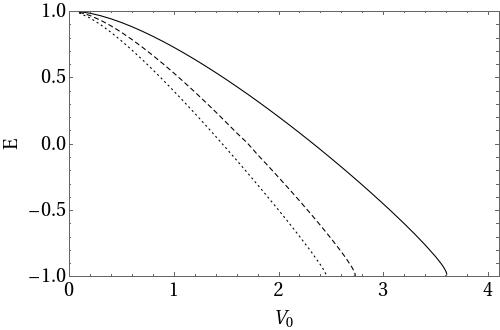}
\caption{Bound-state spectrum for $a=0.5$ and three different values of $x_0$. Solid line represents $x_0=0$, dashed-line represents $x_0=-0.5$, and dotted line represents $x_0=-1$.}
\label{EvsV0_all}
\end{figure}

 Antiparticle bound states appear in all  cases considered. For $a=0.5$, we moved the width parameter $x_0$ from $0$ to $-5$.  Fig \ref{EvsV0_all} shows the bound-state spectrum for $a=0.5$ and three different values of $x_0$, we can observed that the value of $V_\textnormal{cr}$ decreases as $\left|x_0\right|$ increases.  Note that for $x_0=0$ we  recover the results for the cusp potential well \cite{rojas:2006b}, also if we consider $x_0 \neq 0$ and $a \rightarrow 0$ we recover the result for the square potential well \cite{greiner:1987}.  
 
\begin{figure}[th!]
\centering
\includegraphics[scale=0.5]{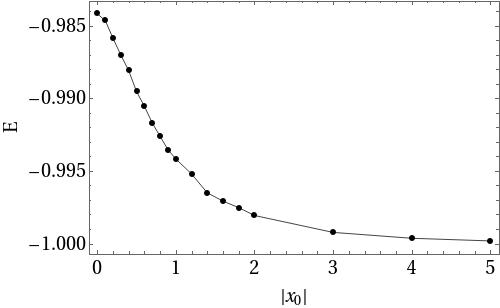}
\caption{Energy vs width of the smooth potential well for $a=0.5$.}
\label{Evsx0}
\end{figure}

 Fig. \ref{Evsx0} shows the behavior of  the energy $E$ versus the potential parameter $x_0$. We can observe that as the value of $\left|x_0\right|$ increases, the energy value, for which antiparticle states appears, decreases.

\vspace{1cm}

\section{Conclusions}

In this paper we have studied the bound state solutions of the Klein-Gordon equation in presence of a smooth potential well. We have calculated the energy eigenvalues as function of the parameters of the potential, and the antiparticle bound state appears in all cases considered. For certain values of the smoothness and width of the smooth potential well we recover the results of the cusp potential well \cite{rojas:2006b} and the square potential well \cite{greiner:1987}.  

\bibliographystyle{unsrt}
\bibliography{well_cusp}

\end{document}